\documentclass[sigconf,10pt,nonacm]{acmart}
\title{Can LLMs Forecast Internet Traffic from Social Media?}

\usepackage{subcaption}
\usepackage{enumitem}
\usepackage{tikz}
\newcommand*\circled[1]{\tikz[baseline=(char.base)]{
            \node[shape=circle,draw,inner sep=1pt] (char) {#1};}}

\author{Jonatan Langlet}
\affiliation{%
  \institution{KTH Royal Institute of Technology\\\& Digital Futures}%
  \city{Stockholm}
  \country{Sweden}%
}
\author{Mariano Scazzariello}
\affiliation{%
  \institution{RISE Research Institutes of Sweden}%
  \city{Stockholm}
  \country{Sweden}%
}
\author{Flavio Luciani}
\affiliation{%
  \institution{Namex}%
  \city{Rome}
  \country{Italy}%
}
\author{Marta Burocchi}
\affiliation{%
  \institution{Namex}%
  \city{Rome}
  \country{Italy}%
}
\author{Dejan Kostić}
\affiliation{%
  \institution{KTH Royal Institute of Technology}%
  \city{Stockholm}
  \country{Sweden}%
}
\author{Marco Chiesa}
\affiliation{%
  \institution{KTH Royal Institute of Technology}%
  \city{Stockholm}
  \country{Sweden}%
}

\newcommand{\smartparagraph}[1]{\vspace{.05in}\noindent\textbf{#1}}
\makeatletter
\newcommand{\DeclareLatinAbbrev}[2]{%
  \DeclareRobustCommand{#1}{%
    \@ifnextchar{.}{\textit{#2}}{%
      \@ifnextchar{,}{\textit{#2.}}{%
        \@ifnextchar{!}{\textit{#2.}}{%
          \@ifnextchar{?}{\textit{#2.}}{%
            \@ifnextchar{)}{\textit{#2.}}{%
              {\textit{#2.,\ }}}}}}}}%
}
\makeatother
\DeclareLatinAbbrev{\eg}{e.g}
\DeclareLatinAbbrev{\Eg}{E.g}
\DeclareLatinAbbrev{\ie}{i.e}
\DeclareLatinAbbrev{\Ie}{I.e}
\DeclareLatinAbbrev{\etc}{etc}
\DeclareLatinAbbrev{\etal}{et~al}

\begin{abstract}
    Societal events shape the Internet's behavior. 
The death of a prominent public figure, a software launch, or a major sports match can trigger sudden demand surges that overwhelm peering points and content delivery networks. 
Although these events fall outside regular traffic patterns, forecasting systems still rely solely on those patterns and therefore miss these critical anomalies.

Thus, we argue for \emph{socio-technical systems} that supplement technical measurements with an active \textit{understanding of the underlying drivers}, including how events and collective behavior shape digital demands.
We propose traffic forecasting using signals from public discourse, such as headlines, forums, and social media, as early demand indicators.

To validate our intuition, we present a proof-of-concept system that autonomously scrapes online discussions, infers real-world events, clusters and enriches them semantically, and correlates them with traffic measurements at a major Internet Exchange Point.
This prototype predicted between 56-92\% of society-driven traffic spikes after scraping a moderate amount of online discussions.

We believe this approach opens new research opportunities in cross-domain forecasting, scheduling, demand anticipation, and society-informed decision making.

\end{abstract}

\begin{document}

\maketitle

\section{Introduction}\label{sec:introduction}
    With the exponential growth of Internet traffic~\cite{shahraki2021comprehensive,cisco_internet_report} driven by IoT devices, high-resolution streaming, and cloud services, accurate Internet traffic forecasting has become more crucial than ever to prevent costly service outages, degraded user experience, and inefficient resource usage~\cite{ferreira2023forecasting}.
    
    Today's operators forecast utilization patterns, typically based on \textit{historical trends}, to optimize their infrastructures~\cite{bega2019deepcog}.
    However, such forecasting techniques often fall short in anticipating large statistical anomalies, such as extreme sudden surges in utilization caused by impactful non-recurring events like game releases, tournaments, TV series, sports broadcasts, sociopolitical\slash cultural collective attention, and more~\cite{zhao2021event,feldmann2020view,netflixTysonPaulDown_techcrunch,pimpinella2022using,flavio_elephanteffect}. 
    
\begin{figure}[t]
        \centering
        \includegraphics[width=1.0\linewidth]{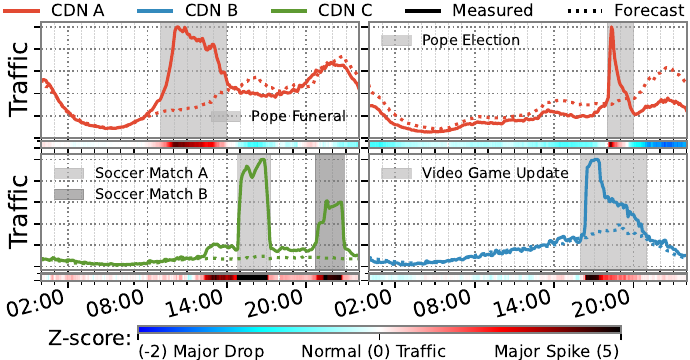}
        \caption{Traffic forecasters cannot predict pattern-breaking event-driven spikes. Each plot shows measured traffic (solid) versus statistical forecasts (dotted) for major CDNs. Z-score color bands indicate the magnitude of deviation. Note how spikes align closely with societal events. Scales and identities omitted per confidentiality policy. }
        \label{fig:motiv_traffic_spikes}
    \end{figure}

    Fig.~\ref{fig:motiv_traffic_spikes} illustrates four traffic spikes: one video game update (bottom right) and three spikes triggered by the death of the Pope in April 2025: the broadcast of the funeral (top left), the broadcast of the papal election (top right), and two last-minute postponed soccer matches (bottom left). %
    These forecasting failures, stemming from the non-recurring and unexpected nature of the events, can result in tangible operational disruptions that directly impact the end-user experience and system reliability~\cite{netflixTysonPaulDown_techcrunch,flavio_elephanteffect}. 

    \looseness=-1\fussy
    Operators often resort to \textit{manual} adjustments to prepare their infrastructure for such events. However, \textit{proactively} discovering these events is a time-consuming and complex task. In many cases, operators depend on informal communication channels, such as peer discussions or community chatter, to become aware of potentially impactful events. 
    This reliance on ad hoc knowledge sharing has led to recent initiatives in which operators manually exchange forecasts of anticipated traffic peaks~\cite{cdnalliance-traffic-radar}. 
    Even if an event is known, inferring its impact still demands \textit{extensive} manual research into audience size, timing, regionality, and expected data footprint~\cite{flavio_elephanteffect}. 
    This entire process is highly resource-intensive, even for hyperscalers, prompting operators to resort to broad overprovisioning as a safeguard~\cite{bega2019deepcog}.
    However, overprovisioning imposes increased operational costs and energy consumption.
    Moreover, as confirmed through conversations with operators, explaining the root societal cause of a spike is often not straightforward, even after a significant spike is verified.
    This underscores the need for \emph{predictive} \emph{socio-aware} technologies that understand societal dynamics and their impact on infrastructure.
    
    Although predicting these event-driven spikes is operationally critical, no existing system can reliably extract, structure, and quantify such events, and estimate their digital impact~\cite{zhao2021event}. 
    Today's forecasting systems are largely grounded in low-level, system-centric signals \eg CPU utilization~\cite{janardhanan2017cpu}, historical traffic traces~\cite{ramakrishnan2018network}, and CDN cache hit rates~\cite{berger2018towards}.

    \smartparagraph{Bridging society and infrastructure.} 
    We argue that \textit{now is the time} for systems to stop treating infrastructure as isolated from society, and begin modelling the reality that drives digital demands. 
    Large Language Models (LLMs)~\cite{vaswani2017attention} represent a \textit{transformative opportunity}: their unprecedented ability to \emph{interpret} public discourse, \emph{understand} cultural context, and \emph{anticipate} collective behavior makes them uniquely suited to infer relevant events, estimate regional hype, and inform of their infrastructural impact.
    This shift enables the development of insightful control systems, which we call \emph{socio-technical systems}, that will adapt to societal dynamics, and even go beyond them. For example, socio-technical systems may need to reason both about domain-specific patterns, \eg whether a soccer team's qualification status affects audience interest in upcoming matches, or reason about synchronized player activities in multiplayer games~\cite{eve-outage}.

    Conversely, these systems could allow infrastructure to influence society. For instance, they may identify events competing over the same resources, raising the question: Should impactful events be scheduled to avoid overlap, mitigating the risk of degraded user experience due to network congestion?
    What about when \emph{event audiences} overlap?

    \smartparagraph{Decoding traffic spikes via public discourse. }
    In this paper, we show how Transformers and LLMs can predict traffic spikes by understanding the underlying societal drivers. 
    We demonstrate the untapped potential in analyzing public discussions, showing that AI can effectively \emph{uncover} and \emph{track} the most relevant upcoming events with sufficient precision to inform traffic forecasting systems.
    Our findings raise new technical questions and research directions, such as: How can we identify the CDNs implicitly ``hosting'' an event? And to what extent can we quantify the impact of online engagement by correlating it with shifts in BGP routes?
    
    Our work, combined with experience in operating a major IXP, has produced insights that we believe will accelerate research into socio-technical systems. We hope these findings will spark broader discussion and inspire future work that builds on and expands beyond the scope of this paper.

    \vbox{
    \vspace{0.7em}
    \noindent \textbf{Our main contributions include:}
    \vspace{-0.1em}
    \begin{itemize}[leftmargin=*,noitemsep]
        \item A forecast system for network traffic anomalies using societal signals extracted from public discourse.
        \item A proof-of-concept AI pipeline that extracts, enriches, deduplicates, clusters, and tracks events to produce abstractions optimized for spike prediction.
        \item A demonstration that unstructured chatter in aggregate conveys enough information to infer traffic-driving events.
        \item A public dataset with thousands of traffic-driving events.
        \item A discussion of opportunities and challenges enabled by socio-technical network systems.
    \end{itemize}
    }

\begin{figure}[t]
    \centering
    \includegraphics[width=\linewidth]{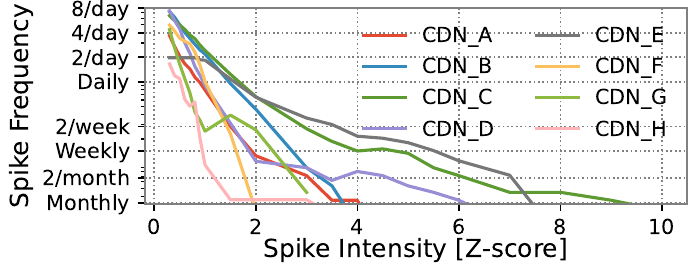}
    \caption{Significant traffic spikes are common, most of which driven by real-world events. We show the spike frequency for a set of major (anonymized) CDNs. The frequency is presented for a range of spike intensities.}
    \label{fig:motiv_num_spikes}
\end{figure}
\section{Motivation \& Background}\label{sec:motivation}
    Real-world events frequently trigger sharp, unexpected traffic surges~\cite{tyson-surge,wwe-surge,fortnite-surge,flavio_elephanteffect}, leading to degraded user experience~\cite{flavio_elephanteffect} or even complete service unavailability~\cite{netflixTysonPaulDown_techcrunch}. 
    Fig.~\ref{fig:motiv_traffic_spikes} illustrates this phenomenon using four real-world day-CDN pairs, comparing measured IXP traffic against baseline forecasts for major CDNs serving a European capital.
    These traffic spikes are driven by societal events, \ie the papal funeral and election, as well as regionally important soccer matches. 
    Such spikes are directly linked to real-world events, and their characteristics are further shaped by additional societal and geographical factors. For instance, the two soccer matches exhibit distinct traffic patterns: the afternoon match ($\sim$3pm) involves a local city team, driving more interest, whereas the evening match ($\sim$9pm) is between non-local teams, resulting in still noticeable but less engagement.

    \newpage
    \smartparagraph{Traffic forecasting lacks real-world context.}
    Despite their operational importance, event-driven traffic surges are routinely missed by state-of-the-art forecasting models~\cite{9838262,saha2022empirical,ferreira2023forecasting}. These models, trained solely on historical patterns, are fundamentally unaware of one-off societal events. This context blindness has \textit{tangible} consequences for end users. For instance, Netflix experienced a high-profile outage during the Tyson–Paul boxing match~\cite{tyson-surge}, resulting in over a million outage reports from 50 countries~\cite{netflixTysonPaulDown_techcrunch}.
    These incidents underscore the critical need for a systematic method to bridge the gap between real-world event dynamics and network traffic patterns, reducing the dependency on costly trial-and-error adjustments.

    \smartparagraph{Statistical forecasting often misses critical spikes.}
    Fig.~\ref{fig:motiv_num_spikes} shows that traffic spikes of varying intensities occur regularly at most CDNs.
    Minor fluctuations (\eg Z-scores~\cite{zscore} below 2) are common and typically harmless.
    In contrast, major spikes capable of overwhelming the infrastructure appear multiple times per year, particularly for general-purpose CDNs with limited insight into the content they serve.
    These are precisely the moments when accurate forecasting matters most and where traditional, history-based models tend to fail. While no public datasets link societal events to traffic surges, our manual investigation of real-world traffic measurements at a major IXP (see Sec.~\ref{sec:evaluation_coverage}) suggests that most major spikes are, in fact, \textit{driven by real-world events}.

    \smartparagraph{Predicting traffic requires more than a calendar: it needs cultural context.}
    Previous studies have integrated real-world events into network traffic prediction~\cite{pimpinella2022using,10633573,9424560,wang2024news}, but they either assume events are already pre-detected and structured~\cite{pimpinella2022using,10633573,9424560} or fail to \emph{track} their dynamics~\cite {wang2024news}, limiting effectiveness as events evolve.
    Yet, anticipating events is not enough, \textit{unexpected changes} in event timing or nature can shift traffic in ways that static models fail to predict (\eg the postponed matches in Fig.~\ref{fig:motiv_traffic_spikes}). Operators must understand \emph{how much} traffic will flow, \emph{where} and \emph{when}.
    Take the two matches: although both were part of the same tournament, played on the same day, and delivered through the same CDN, their traffic signatures differ significantly. As noted, this divergence stems from underlying societal factors, \eg fanbase engagement, that shape demand.
    
    Unfortunately, such contextual metadata is not readily available through traditional event scraping. They are highly dependent on region-specific social knowledge \eg which sports are popular in a given country or what carries cultural significance. 
    For example, while a soccer match may lead to surges at a European ISP, a US-based ISP is as likely to observe them during NBA games.
    Capturing this diversity requires not only semantic understanding of events, but also awareness of the broader cultural and societal context in which they occur. 

    \smartparagraph{Online chatter holds the missing signal.}
    This is where online discussions become invaluable. Platforms like Reddit, X\slash Twitter, and news aggregators often hint at traffic-driving events well in advance, signaling timing, scale, sentiment, and availability, either explicitly or implicitly. Unlike static calendars, these dynamic sources reflect what people are \textit{actually planning to do}. In aggregate, they provide rich insight into public interest and emerging trends that, if parsed and structured, could form a live, evolving feed of future digital demands. 
    Even in the papal election case of Fig.~\ref{fig:motiv_traffic_spikes}, where timing is unpredictable due to the conclave’s uncertain duration, the system can detect rising interest online and anticipate a spike as soon as the new pope is announced.

    \smartparagraph{LLMs are key enablers.} Extracting meaningful information from unstructured and diverse sources is a known complex challenge~\cite{STIEGLITZ2018156}. 
    In recent years, LLMs have shown exceptional performance across various domains, thanks to their deep semantic and contextual awareness~\cite{zhao2025surveylargelanguagemodels}. They excel at filtering out irrelevant or noisy data~\cite{gur2023understandinghtmllargelanguage}, and can reason through intricate relationships between topics or concepts~\cite{bubeck2023sparksartificialgeneralintelligence}.
    While their internal knowledge is limited by a training cutoff, this constraint can be lifted through techniques like tool-augmented training~\cite{shen2024llm}, where models dynamically use search tools, and Retrieval-Augmented Generation (RAG)~\cite{gao2023retrieval}, where models access real-time external information. 
    We believe that, when combined with external knowledge sources, LLMs are uniquely equipped to infer relevant events from diverse, unstructured data, placing them in the proper context, both temporally and geographically, while understanding the underlying societal and cultural forces.

    \smartparagraph{We argue for society-aware forecasting.}
    We propose an AI pipeline that extracts, understands, and clusters societal events from online chatter and uses them to predict event-driven traffic. This approach enables proactive allocation, planning, and reduced operational cost. 
    Beyond immediate application, it also opens a new line of research: inferring digital behavior from societal signals at population-scale.\footnote{We discuss ethical implications of population-scale ``surveillance'' in Sec.~\ref{sec:discussion}.}

\section{Proof of Concept}\label{sec:design}
    \begin{figure}[t]
        \centering
        \includegraphics[width=1.0\linewidth]{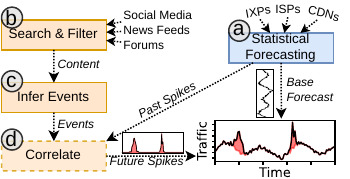}
        \caption{An overview of our approach.}
        \label{fig:system_overview}
    \end{figure}
    To assess whether online discussions can meaningfully inform traffic forecasts, we built a modular pipeline that parses public discourse, extracts latent societal events, and infers their likely digital footprint.
    While our current prototype does not yet close the loop with measured traffic data, it aims to approximate how, where, and when such events might shape future load.
    The overall system is visualized in Fig.~\ref{fig:system_overview}, where each major component is labeled from \circled{a} through \circled{d}.

    \subsection[Context-unaware Forecasting]{Context-unaware Forecasting \circled{a}}\label{sec:design_forecasting}
        We begin by applying standard time series forecasting techniques to model baseline network traffic. These models capture regular patterns such as weekly cycles and seasonal effects, but remain blind to societal context. While more advanced deep learning forecasters exist~\cite{9838262,dl-mobile-traffic}, they too fail to predict event-driven spikes due to this lack of context. Since our goal is not to optimize time series accuracy but to isolate and understand unexpected load surges, a simple rolling statistical approach suffices as they are known for their reliability~\cite{saha2022empirical}.
        In our solution, by subtracting the predicted baseline from observed traffic, we isolate residual spikes that may reflect external influences or natural variability. Based on manual inspection, nearly all statistically significant spikes (for example, those with $Z \geq 3$) appear to correspond to real-world events (see Sec.~\ref{sec:evaluation_coverage}).
        This decomposition yields a collection of historical anomalies that form the foundation for identifying event-driven patterns and learning how such events may drive future traffic spikes.

    \subsection[Online Data Collection]{Online Data Collection \circled{b}}\label{sec:design_collection}
        To support event inference, we collect and organize public discourse from Reddit, selected for its open API, topical breadth, and active user base. Posts are filtered using combinations of search terms, communities, engagement thresholds, and the presence of outbound links. For each selected post, we retrieve its content, top-level comments, and any linked webpages to capture a rich context.
        The resulting material is automatically cleaned, preprocessed, and compiled into a single content record per discussion thread.

    \subsection[Event Inference]{Event Inference \circled{c}}\label{sec:design_inference}
        \begin{table}[tbp]
            \centering
            \resizebox{\columnwidth}{!}{%
            \begin{tabular}{@{}lll@{}}
                \toprule
                \textbf{Description} & \textbf{Data Type} & \textbf{Aggregation Method} \\ 
                \midrule
                Date of the event & String & Fixed on creation \\
                Time of the event & String & Fixed on creation \\
                Short description of the event & String & Fixed on creation \\
                Event Category & String & Plurality string \\
                List of relevant entities & String (List) & Entries w/ $\geq 2$ votes \\
                Platforms and services & String (List) & Entries w/ $\geq 2$ votes \\
                Internet data per user & Integer & Median value \\
                Estimated global audience size & Integer & Median value \\
                Relevance across continents & Float (List) & Per-entry median values \\
                Relevance across nations & Float (List) & Per-entry median values \\
                Duration of the traffic spike & Float & Median value \\
                Likelihood to happen as described & Integer (0–10) & Median value \\ 
                Semantic categorization vector & Integer (List) & Multi-level clustering \\ 
                \bottomrule
            \end{tabular}
            }
            \caption{Overview of Event Metadata.}
            \label{tab:event_metadata}
        \end{table}
        The system converts each content record into a streamlined text format suitable for LLM processing. A reasoning model (\ie 70B DeepSeek-R1~\cite{llama70}) identifies all upcoming events that might influence network traffic, returning three core attributes: headline, date, and time.\\
        This minimal scope avoids coalescing multiple events within a single discussion, which is common in posts about fixtures, tours, and similar sources with multiple announcements.
        The extracted events are parsed into structured abstractions and written into the database for downstream processing.

        \smartparagraph{Metadata Inference.}
        Each extracted event is annotated with a rich set of metadata (see Table~\ref{tab:event_metadata}), inferred one at a time by an ensemble of reasoning models \mbox{(\ie 14B DeepSeek-R1~\cite{qwen14})}. 
        For each field, the system aggregates predictions from three independent LLM runs, repeating if consensus fails. 

        \noindent
        The core initial metadata, that is, the event category and associated entities (\eg video game franchises or sport teams), are inferred directly from the extractor output and content record.
        Later stages use RAG, incorporating relevant Wikipedia articles to provide context and improve inference quality across attributes, such as audience size, geographic reach, and expected data usage.

        \smartparagraph{De-duplication.}
        To handle redundancy, our prototype computes embeddings for all events using an encoder-only Transformer~\cite{encoder_only} applied to their free-text summaries. Pairwise cosine similarities of the embeddings are used to identify semantically highly similar events occurring on the same date, which are treated as duplicates.
        When duplicates are found, their content records and metadata inferences are re-linked under a single event abstraction, and the redundant entries are removed.
        All metadata is then re-inferred using the expanded context available after merging.

        \smartparagraph{Semantic Categorization.}
        Since LLM-predicted traffic impacts are inherently noisy and lack ground-truth information, we must correlate events to traffic using historical patterns.
        Exact matches between past and future events are rare, so we instead learn from semantically similar events. For example, predicting the impact of a UEFA semi-final benefits from prior data on sports matches in general, past UEFA games, and matches involving the same teams (\eg due to fan bases and regional importance).
        
        Our prototype clusters event embeddings at multiple levels of granularity using k-means, ranging from broad groupings like ``sports'' at $k=10$ to ``soccer semi-finals involving team X'' at $k=10,000$.
        Each event is assigned to a cluster at each level, forming a semantic signature that situates it among related events. These dynamically learned categories can be used as features in a prediction model, letting it learn from prior examples even when exact matches do not exist.

         \vspace{-0.9em}
        \subsection[Spike-Event Correlation]{Spike-Event Correlation \circled{d}}
        \vspace{-0.2em}
        Our event abstractions are designed to integrate with context-aware time-series forecasters.
        While an optimal solution remains an open research challenge, a natural approach is to train models such as the Temporal Fusion Transformer on historical traffic and structured event metadata.
        Each event contributes a rich set of predictive features, including timing, expected data intensity, duration, geographic scope, relevant platforms, and the multi-level semantic categorization vector. These features help the model learn how different types of events affect different networks in different regions.
        
        By modeling a local time window around each event (\eg three days), the forecaster can capture both immediate spikes and adjacent effects. Over time, it may also learn negative correlations, such as cases where attention to one event draws users away from other platforms.

    \vspace{-0.9em}
    \subsection{Future Work}\label{sec:futurework}
        \vspace{-0.2em}
        While our prototype suffices to demonstrate the feasibility of using online chatter to explain and anticipate traffic spikes, we believe the underlying idea enables more powerful spike prediction systems beyond this initial proof of concept.

        \vspace{-0.2em}
        \smartparagraph{Global semantic context.}
        The prototype successfully abstracts events from online content with both direct and implied references.
        While most relevant context is often captured in surrounding discussions, some broader influences may go unmentioned yet still affect network behavior. For example, an ongoing papal election may significantly alter engagement with a newly released film about the conclave, even if that connection is not made explicit in online chatter.

        Incorporating a more global or cultural context, such as prevailing themes or public interests (\eg a surge in the ``zombie'' genre), could enhance inference quality. Although our current semantic categorization supports some thematic generalization and spike tuning, it lacks semantic reasoning around broad cultural signals and the current zeitgeist.

        \vspace{-0.2em}
        \smartparagraph{Inferring impacted networks.} Events rarely specify which networks will serve content, and their delivery paths could dynamically shift. For example, live sports broadcasts may change providers or underlying CDNs~\cite{10.1145/2785956.2787500,cbs_serie_a,fox_serie_a}. 
        Moreover, routing paths to a given content provider can vary over time due to shifting routing policies or unexpected outages~\cite{baltic-cable}. These route changes can significantly affect the traffic load at different locations depending on the time.
        
        To address this, one solution is to infer likely target networks through learned event-to-traffic patterns and known broadcaster relationships. 
        This is challenging, as commercial agreements are rarely disclosed. One possible approach is to identify events associated with the same inferred service and compare their observed traffic spikes against expected patterns. Consistent deviations can reveal which CDN likely serves the content, thereby tracking these changes.
        Also, one could incorporate knowledge about BGP updates to infer likely impacted on-path networks (\eg IXPs).

        \smartparagraph{Filtering massive online content.} Reddit receives nearly a million new posts daily~\cite{reddit_stats}, whereas X\slash Twitter sees hundreds of millions~\cite{twitter_stats}. Only a small fraction relates to potentially traffic-driving events, making large-scale analysis costly. To reduce overhead, this stream of content must be filtered before analysis to increase the density of relevant signals and improve the timeliness spike prediction. However, overly aggressive filtering risks missing events with low global visibility but high local impact, such as regional sports matches.

        One could design a feedback mechanism that continuously trains a content filtering model based on what content typically leads to high-quality event information. For instance, an encoder-only Transformer can be fine-tuned to quickly estimate contents' relevance before event inference.
        
\section{Feasibility Evaluation}\label{sec:evaluation}
    To the best of our knowledge, \emph{tracking} events, sentiment, and Internet utilization through online chatter is entirely novel.
    Existing systems do not track the societal context behind anomalies, whereas our entire focus is on doing so.
    As such, this section does not attempt to compare against prior work but instead examines whether online discussions carry enough signal to support event-driven forecasting.

    For our proof-of-concept, we scraped 38,000 webpages, including Reddit, linked websites, and Wikipedia.
    From this, $\sim$10,000 unique traffic-driving events were inferred and enriched through $\sim$200 A100~\cite{a100} compute hours.

    Events will be released as a dataset alongside this paper.

    \subsection{Spike Coverage}\label{sec:evaluation_coverage}
        Unfortunately, there are no publicly available datasets mapping societal events to their network impact.
        Therefore, we manually curated a small set of traffic spikes with known causes, identified through public sources such as news articles, release calendars, and operator blog posts. Due to the extensive effort in manually researching individual spikes, we settled with 29 verified event-spike pairs\footnote{We selected spikes with contiguous $Z\geq2$ and duration $\geq20$ minutes from between March and July 2025.}.

\begin{figure}[t]
            \centering
            \includegraphics[width=\linewidth]{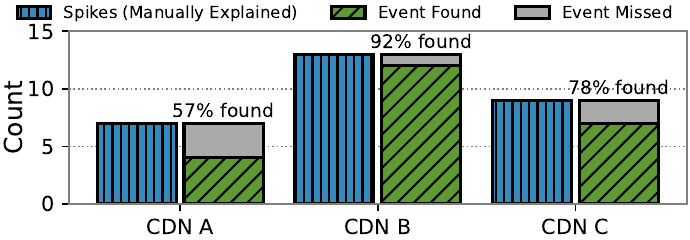}
            \vspace{-2.5em}
            \caption{Fraction of manually investigated traffic spikes that were successfully predicted in our prototype, across three CDNs.}
            \label{fig:eval_spike_coverage}
            \vspace{-1.5em}
        \end{figure}

        \looseness=-1
        For each explained spike, we checked whether our system had inferred the same event at that time. \\
        As shown in Fig.~\ref{fig:eval_spike_coverage}, the system correctly matched a majority of these cases, with coverage rates ranging from 57\% to 92\% across CDNs. 

        Although these results already support our idea, some events are missed. These fall into one of two categories: (1) the event \emph{is} discussed in advance, but the scraper had not yet crawled that content, or (2) the event is spontaneous, triggered by breaking news, and therefore appeared with minimal notice (\eg the death of a public figure).

        \vspace{-0.2em}
        \smartparagraph{Takeaway:} Public online discourse holds great promise as a source for accurately predicting real-world traffic events, even without manual tuning or ground-truth labels.

    \vspace{-0.7em}
    \subsection{Event-discussion Timelines}\label{sec:evaluation_timeline}
\begin{figure}[t]
            \centering
            \includegraphics[width=\linewidth]{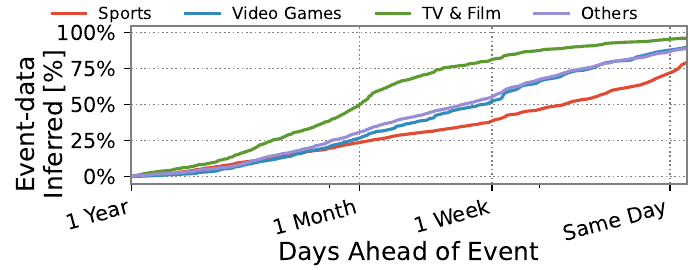}
            \vspace{-2.0em}
            \caption{Cumulative fraction of event mentions by time-to-event, grouped by category. Many events are discussed well in advance, while context and engagement often intensify closer to the event date.}
            \label{fig:eval_advance_notice}
            \vspace{-1.9em}
        \end{figure}

        \vspace{-0.3em}
        We extract upcoming events from general Reddit discussions and measure how far in advance they are mentioned. Fig.~\ref{fig:eval_advance_notice} shows the cumulative fraction of events first detected at various lead times, grouped by category. Categories with fewer than 1,000 inferred events are grouped as ``Others''.

        TV \& Film events typically surface early, and nearly half of them are found in Reddit discussions one month before they occur. Sports exhibit shorter notice periods. While fixtures are typically released in a structured format ahead of time, crucial details, like team matchups, recent performances, and audience anticipation, tend to unfold closer to the event date.
        Video game events have a mixed behavior, combining early mentions tied to official release dates with last-minute spikes from reviews or content updates. %

        While this evaluation is based on a prototype pipeline sampling a small subset of online discussions, the trends reflect real structural differences. Some events follow fixed calendars, while others depend on evolving context. For instance, while a movie release might be advertised years in advance, context-dependent information such as actors, budgets, and public interest dynamically evolves up until the release itself. 
        Our results highlight the potential of general-purpose platforms to provide detailed and context-dependent signals for demand-driven events ahead of time.

        \smartparagraph{Takeaway:} Online discussions reveal not just that an event will happen, but when public attention begins to build. Different domains show distinct lead time profiles, reflecting how event-information and public interest evolve.

\vspace{-0.6em}
\section{Discussion}\label{sec:discussion}
    \vspace{-0.6em}
    \smartparagraph{Future application domains.}
    While our primary goal is context-aware traffic forecasting for resource allocation and network management, the event abstraction methodology has applications in many other areas.

    Regional event predictions, for example, can guide \emph{proactive CDN cache placement} in areas of anticipated demand or \emph{anticipate mass human movement} for mobile load balancing, GPS routing, and traffic planning during major events.

    Socio-technical systems are vulnerable to manipulation via misinformation and ``fake news''. However, by systematically correlating predicted events with observed utilization data, these systems could identify and downweight sources whose claims repeatedly fail to materialize in empirical data.

    \noindent
    This weighting offers a novel, empirically grounded signal of source trustworthiness that could extend beyond forecasting to support misinformation detection, media forensics, or trust-aware content ranking in algorithmic feeds.

    \vspace{-0.1em}
    \smartparagraph{Limitations.}
    While this paper highlights the predictive power of online discourse, there are fundamental limitations that constrain the practicality and coverage of this approach.

    First, our method assumes reliable access to public online discussions, either through APIs or web scraping. However, access to such data is neither guaranteed nor stable. APIs may impose strict rate limits, require expensive subscriptions, or be deprecated entirely. Web scraping, while more flexible, is increasingly hindered by anti-bot measures~\cite{cloudflare-ai}. As platforms adopt more aggressive protection against automated access, sustaining large-scale real-time content ingestion becomes more challenging.
    In short, while the Internet talks about what it will do, \textit{it is increasingly hard to listen}.

    Second, certain events occur with little or no advance notice. These include sudden celebrity announcements, geopolitical shocks, and private decision reveals (such as surprise product launches).
    Because our system relies on textual indicators in public platforms to infer future traffic surges, it will either miss such ``spontaneous'' events entirely or only detect them as they occur. In such cases, the predictive power of our system collapses to a real-time explainer at best.

    \smartparagraph{How far should LLMs go?  } 
    As LLMs predict digital events, a key ethical question arises: how far should they go to detect early signals? Some, like traffic spikes from coordinated actions, may stem from private discussions. For example, in games like EVE Online, a 6,000-player assault might only become visible after it happens. Should LLMs ``infiltrate'' such groups for early insight or would that cross an ethical line?

    The situation becomes murkier when game mechanics create fixed, publicly known attack windows. While such structures can make events technically predictable, the actual scale of participation depends on players' strategic choices that are often coordinated in private channels with occasional leaks on public forums. During the largest in-game war~\cite{m2-xfe}, 12,000 players coordinated in private channels an attack during a publicly known attack window. Such private channels make potential outages difficult to anticipate.

    Our prototype avoids non-public data by design, but as LLMs become more persuasive, the line between monitoring and unethical surveillance blurs. Actors must ask not just \textit{can} we predict an event, but \textit{should} we? This goes beyond gaming. Should LLMs predict DDoS attacks if it means learning from semi-private or encrypted spaces like Telegram?

\section{Conclusion}\label{sec:conclusion}

    Societal dynamics and Internet traffic are tightly coupled. Popular events generate surges in demand, while the way discussions unfold provides crucial metadata for anticipating their impact.
    From nearly 10,000 unique events extracted from online discourse, our prototype predicted 57–92\% of major CDN traffic spikes.
    This shows that public chatter already encodes much of tomorrow’s user behavior, and that with the right abstractions, systems can learn to listen.

    We do not offer a complete forecasting solution, but evidence that socio-technical forecasting is both feasible and filled with open questions: how to map events to the delivery networks that serve them, how to capture cultural and regional factors that shape demand, and how to separate fleeting noise from traffic-shaping signals.
    Addressing these challenges will enable infrastructure that is not only reactive but anticipatory of the society that drives it.

\section*{Acknowledgements}
    This work has been partially supported by Knut and Alice Wallenberg Foundation (Wallenberg Scholar Grant for Prof. Dejan Kostić), Vinnova (Sweden's Innovation Agency), the Swedish Research Council (agreement No. 2021-04212), and KTH Digital Futures.
    The computations were enabled by resources provided by the National Academic Infrastructure for Supercomputing in Sweden (NAISS) through grant agreements no. 2024/22-938 \& 2025/22-924.

\bibliographystyle{ACM-Reference-Format} 
\bibliography{references}

\end{document}